\documentclass[aps,pra,twocolumn,8pt,showpacs]{revtex4}
\usepackage{mathrsfs}
\usepackage{graphicx}
\usepackage{amsmath}
\usepackage{amssymb}

\usepackage{hyperref}
\usepackage{subfigure}

\usepackage{graphicx}
\usepackage{dcolumn}
\usepackage{bm}
\usepackage{amsmath}
\usepackage{amssymb}
\usepackage{latexsym}
\usepackage{epsfig}
\usepackage{amsbsy}
\usepackage{array}
\usepackage{amssymb}
\usepackage{setspace}
\usepackage{bm}
\usepackage{epstopdf}
\usepackage{subfigure}

\begin{document}

\title{Speeding up adiabatic state conversion in optomechanical system}

\author{Feng-Yang Zhang$^1$\footnote{zhangfy@dlnu.edu.cn}, Wen-Lin Li$^2$, and Yan Xia$^3$\footnote{xia-208@163.com}}
 \affiliation{$^1$School of Physics and Materials Engineering, Dalian Nationalities University, Dalian $116600$, China\\
 $^2$School of Physics and Optoelectronic Technology, Dalian University of Technology, Dalian 116024, China\\
 $^3$Department of Physics, Fuzhou University, Fuzhou 350002, China}

\begin{abstract}
Speeding up adiabatic method has attracted much attention with the wide applications in quantum information processing.
In this paper, two kinds of methods, Lewis-Riesenfeld invariant-based inverse engineering and transitionless quantum driving are applied to implement speeding up adiabatic state conversion in optomechanical system. The perfect population transfer can be achieved within a short time. At last, the energetic cost is analysed for the transitionless quantum driving.
\end{abstract}
\maketitle
\section{Introduction}
Optomechanical system is composed of an optical (or microwave) cavity and a mechanical resonator. And it explores the interaction between light and mechanical motion. This field has made tremendous
progress over the past decades \cite{1}. Optomechanical system has shown the advantage to quantum information processing \cite{2}.

Recently, quantum state conversion between cavity modes of distinctly different wavelengths has been studied \cite{3, 4}.
A major flaw of this scheme that the fidelity is limited by cavity damping, thermal noise in the mechanical
mode, and accuracy of the pump pulses. In order to overcome the effect of
thermal noise on the transfer fidelity, the adiabatic quantum state transfer was proposed \cite{5}. During this process, the quantum states are preserved in a mechanical dark mode with negligible
excitation to the mechanical mode. Simultaneously, Wang and Clerk \cite{6} proposed using a
mechanically dark delocalized mode in a two-cavity optomechanical
system for quantum state transfer, and demonstrated that both intracavity states and itinerantphoton
states can be transferred with high fidelity.

However, an obstacle in
achieving this is the long run time required for the desired
parametric control associated with adiabatic evolution might be problematic, when the decoherence effect is considered.
How to fast achieve and high-fidelity quantum state transfer in optomechanical system is an open question.
There are two kinds of approaches, Lewis-Riesenfeld invariant-based inverse engineering \cite {7, 8, 9, 10, 11} and transitionless quantum driving \cite{12, 13, 14, 15, 16, 17},
are widely applied to speed up adiabatic state conversion.

In this paper, we propose an alternative scheme to speed up adiabatic state conversion in optomechanical system, which is composed of two cavity modes and one mechanical mode. When no damping is considered, we use the method of Lewis-Riesenfeld invariant-based inverse engineering to speed up adiabatic state conversion. When the damping of the system is considered, the approach of transitionless quantum driving is applied to  speed up adiabatic state conversion.
Quantum states can be converted between different
cavity modes by varying the effective optomechanical
couplings. Our scheme has potential applications in quantum information processing.
\section{Model}
We study an optomechanical system, which is composed of two cavity modes and one mechanical mode
coupling via optomechanical forces. After the standard linearization
procedure, under the
rotating-wave approximation, the effective Hamiltonian for this
coupled system is given by ($\hbar=1$) \cite{5, a}
\begin{eqnarray}
H_{s}=\sum_{i=1,2}\Delta_{i}a^{\dag}_{i}a_{i}+g_{i}(a^{\dag}_{i}b+a_{i}b^{\dag})+\omega_{m}b^{\dag}b,
\end{eqnarray}
where $a^{\dag}_{i} (a_{i}) (i=1,2)$ and $b^{\dag} (b)$ are the creation (annihilation) operator for the
\emph{i}th cavity mode and the mechanical
mode, respectively, $\Delta_{i}$ is the laser detuning, $g_{i}$ is the effective linear coupling strength, and $\omega_{m}$ is the mechanical frequency.

We consider the damping of the system. The cavity damping
rates are $\kappa_{i}$ and the mechanical damping rate is $\gamma$. When the condition $\omega_{m}=-\Delta$ is satisfied,
the Langevin equation in the interaction picture can be
written as
\begin{eqnarray}
i\frac{d}{dt}A(t)=\mathcal{N}A(t),
\end{eqnarray}
where the parameters $A(t)=[a_{1}(t),~b(t),~a_{2}(t)]^{T}$
and the dynamic matrix can be written as
\begin{eqnarray}
\mathcal{N}=\left(
                 \begin{array}{ccc}
                   -i\frac{\kappa_{1}}{2} & g_{1} & 0 \\
                   g_{1} & -i\frac{\gamma}{2} & g_{2} \\
                   0 & g_{2} & -i\frac{\kappa_{2}}{2} \\
                 \end{array}
               \right). \label{b}
\end{eqnarray}

\section{Adiabatic state conversion}
For the simple case of zero dampings $\kappa_{a}=\kappa_{b}=\kappa_{c}=0$, the eigenvalues of the dynamic matrix are
$E_{1}=0$, $E_{2}=g_{0}/\sqrt{2}$, and $E_{3}=-g_{0}/\sqrt{2}$ with the $g_{0}=\sqrt{g_{1}^{2}+g_{2}^{2}}$. The eigenvalue $E_{1}=0$ corresponding to eigenstate is $\psi_{1}=[-g_{2}/g_{0},0, g_{1}/g_{0}]^{T}$, which is a mechanical
dark mode that only involves the cavity modes. The population transfer of the cavity modes is realized by the dark state. The quantum state to be transferred is initially stored in cavity mode $a_{1}$. Then, $g_{1}$ is adiabatically decreasing and $g_{2}$ is adiabatically increasing. The information is transferred from cavity mode $a_{1}$ to mode $a_{2}$.

As an example, we choose the time-dependence coupling
strengths, which are expressed by \cite{d, e}
\begin{eqnarray}
g_{2}(t)=Gf(t-\tau); g_{1}(t)=Gf(t); \nonumber
\end{eqnarray}
\begin{eqnarray}
f(t)=\{\begin{array}{c}
       \sin^{4}(\pi t/T)~~~(0<t<T) \\
       0            ~~~~~(otherwise),
     \end{array}
\end{eqnarray}
where $G$ is the amplitude of the coupling coefficients, $\tau$ represents the deviation of the
time interval between two coupling strengths, and $T$ is the period. Figure 1(a) shows the evolution of coupling coefficients $g_{1}$ and $g_{2}$. Figure 1(b) shows the evolution of the populations of the system. The population is not completely transferred from cavity mode $a_{1}$ to $a_{2}$.

\begin{figure}[t]
\includegraphics*[width=4.2cm, height=2.5cm]{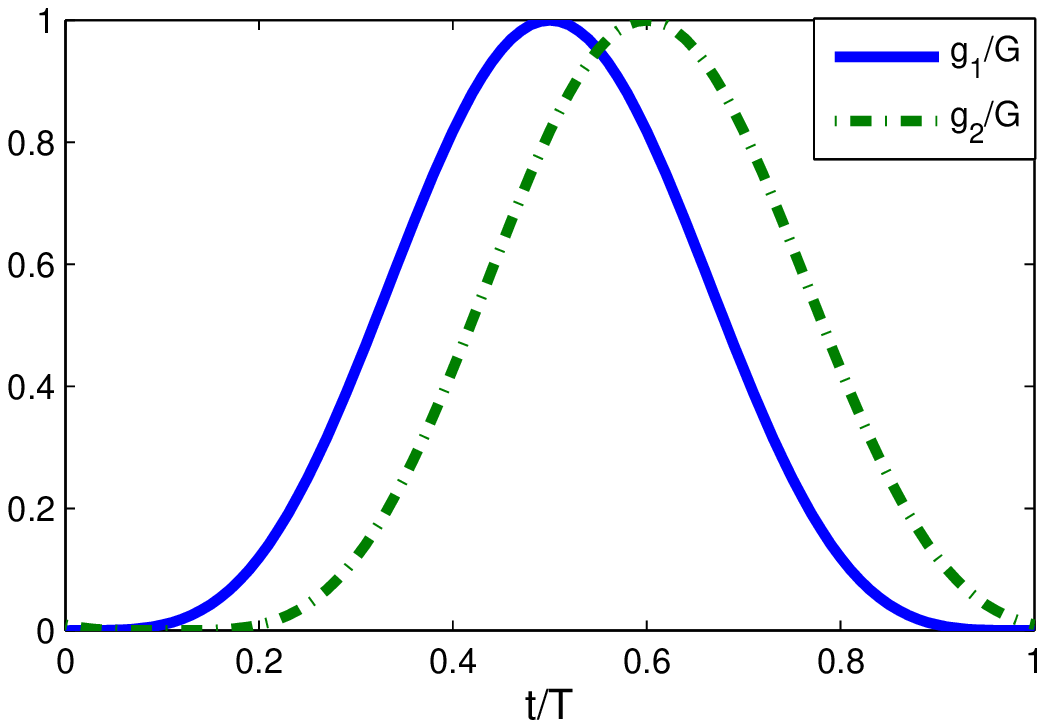} \includegraphics*%
[width=4.2cm, height=2.5cm]{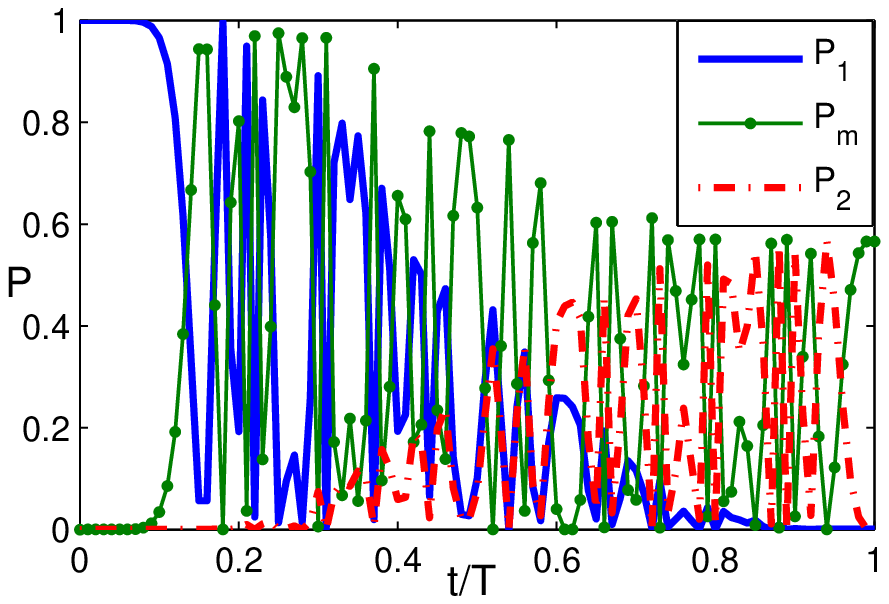}
\caption{(color online) (a) Evolution of coupling coefficients $g_{1}$(solid blue line) and $g_{2}$(dashed green line) with $\tau=0.1T$. (b) Simulation of the quantum state populations transfer of the cavity mode $a_{1}$ (solid blue line), cavity mode $a_{2}$ (dashed red line), and  mechanical mode $b$ (solid-dot green line) with $G=10^3$.}
\end{figure}

\section{Shortcut to adiabatic passage.}
Converting quantum states between information memory units can have profound
influences on quantum information processing. The adiabatic methods have been proposed for different physical systems.
An obstacle in achieving this is the long run time required for the desired parametric
control associated with adiabatic evolution. Here, we discuss how to implement the ultrafast converting quantum states.

When the no damping is considered, the Eq. (\ref{b}) possesses SU(2) dynamical symmetry. To speed up the converting quantum states,
we introduce an invariant $I(t)$, which satisfies $\partial I(t)/\partial t=-i[I(t), \mathcal{N}]$ and is given by \cite{8, 9, 21}
\begin{equation}
I(t)=\Omega\left(
\begin{array}{ccc}
       0 & \cos\alpha\sin\beta & -i\sin\alpha \\
       \cos\alpha\sin\beta & 0 & \cos\alpha\cos\beta \\
       i\sin\alpha & \cos\alpha\cos\beta & 0
     \end{array}
     \right),
\end{equation}
where $\Omega$ is an arbitrary constant with units of frequency to keep $I(t)$ with dimensions of energy, and the time-dependent
auxiliary parameters $\alpha$ and $\beta$ satisfy the relations $\dot{\alpha}=g_{1}\cos\beta-g_{2}\sin\beta$ and $\dot{\beta}=\tan\alpha(g_{2}\cos\beta+g_{1}\sin\beta)$. Thus, the coupling coefficients also can be expressed by parameters $\alpha$ and $\beta$ as follows: $g_{1}=\dot{\beta}\cot\alpha\sin\beta+\dot{\alpha}\cos\beta$ and
$g_{2}=\dot{\beta}\cot\alpha\cos\beta-\dot{\alpha}\sin\beta$. The eigenvalues of the invariant $I(t)$ are $\epsilon_{1}=0$, $\epsilon_{2}=-1$, and $\epsilon_{3}=1$ corresponding to the eigenstates $\varphi_{1}=[\cos\alpha\cos\beta, ~-i\sin\alpha, ~-\cos\alpha\sin\beta]^{T}$,
 $\varphi_{2}=\frac{1}{\sqrt{2}}[\sin\alpha\cos\beta-i\sin\beta, ~i\cos\alpha, ~-\sin\alpha\sin\beta-i\cos\beta]^{T}$, and $\varphi_{3}=\frac{1}{\sqrt{2}}[\sin\alpha\cos\beta+i\sin\beta, ~i\cos\alpha, ~-\sin\alpha\sin\beta+i\cos\beta]^{T}$, respectively. The converting quantum states between information memory units along the invariant eigenstate $\varphi_{1}$ in a given time $T$, which is the total evolution time. Therefore, we choose the parameters for $\alpha$ and $\beta$ as follows: $\alpha=\xi$ and $\beta=\pi t/2T$, where $\xi$ is a small value. And we obtain
\begin{eqnarray}
 g_{1}&=&(\pi/2T)\cot\xi\sin(\pi t/2T), \\
 g_{2}&=&(\pi/2T)\cot\xi\cos(\pi t/2T).
\end{eqnarray}
  Figure 2(a) shows the evolution of coupling coefficients
$g_{1}$ and $g_{2}$. Figure 2(b) shows the evolution of the populations of the system. The perfect population
transfer from cavity mode $a_{1}$to $a_{2}$ can be achieved.
\begin{figure}[t]
\includegraphics*[ width=4.2cm, height=2.5cm]{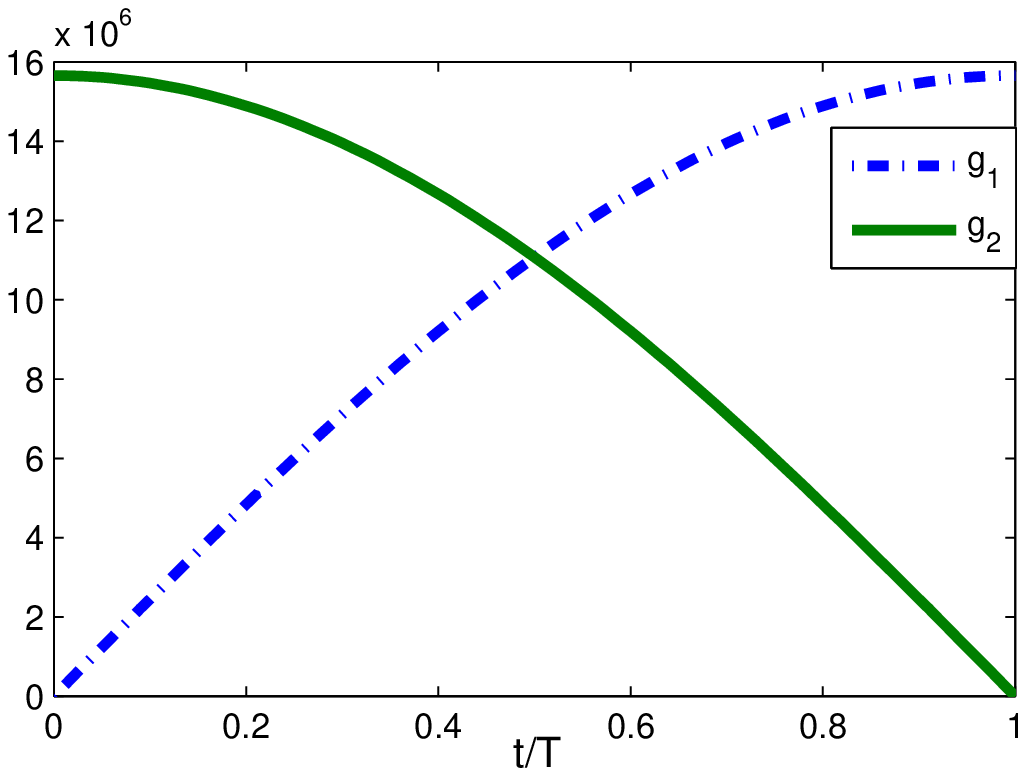} \includegraphics*%
[width=4.2cm, height=2.5cm]{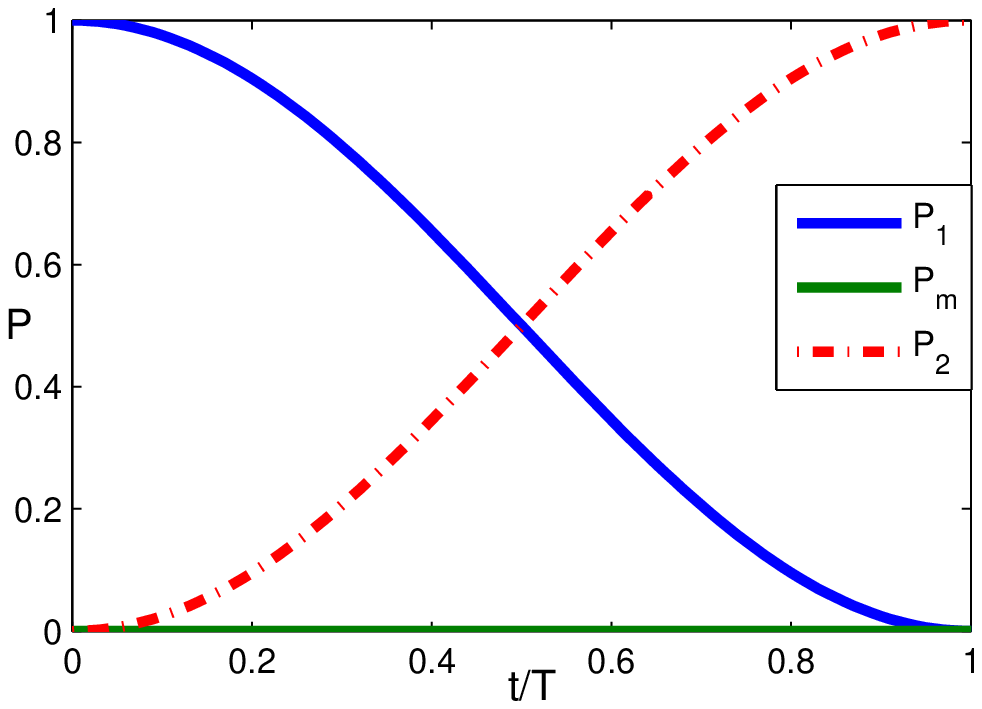}
\caption{(color online) (a) Evolution of coupling coefficients $g_{1}$(solid blue line) and $g_{2}$(dashed green line) with $\xi=0.1, T=1\mu s$. (b) Simulation of the quantum state populations transfer of the cavity mode $a_{1}$ (solid blue line), cavity mode $a_{2}$ (dashed red line), and  mechanical mode $b$ (solid green line).}
\end{figure}

 When the damping of the system is considered. Here, we assume that the cavity modes and the mechanical mode have equal damping rates, i.e., $\kappa_{1}=\kappa_{2}=\gamma=\kappa$. The dynamics matrix (\ref{b}) has the nondegenerate instantaneous eigenstates $|\lambda_{1}(t)\rangle=[g_{1}/g_{0}, 1, g_{2}/g_{0}]^{T}/\sqrt{2}, |\lambda_{2}(t)\rangle=[g_{1}/g_{0}, 1, g_{2}/g_{0}]^{T}/\sqrt{2}$ and $|\lambda_{3}(t)\rangle=[-g_{2}/g_{0},0, g_{1}/g_{0}]^{T}/\sqrt{2}$ with corresponding eigenvalues $E_{1}=g_{0}-i\kappa/2, E_{2}=-g_{0}-i\kappa/2$, and $E_{3}=-i\kappa/2$, respectively.
We introduce a Hamiltonian $\mathcal{H}_{I}(t)$, which steers the dynamics along the
instantaneous eigenstates $|\lambda_{m}(t)\rangle$ without transitions
among them and without phase factors, and it is described by $\mathcal{H}_{I}(t)=i\sum_{m}|\partial_{t}\lambda_{m}(t)\rangle\langle\lambda_{m}(t)|$ which takes the form
\begin{eqnarray}
\mathcal{H}_{I}(t)=
\frac{1}{2}\left(
  \begin{array}{ccc}
    0 & 0 & i\vartheta \\
    0 & 0 & 0 \\
    -i\vartheta & 0 & 0 \\
  \end{array}
\right),\label{c}
\end{eqnarray}
where $\vartheta=(\dot{g}_{1}g_{2}-g_{1}\dot{g}_{2})/g_{0}^{2}$.
This matrix indicates that there should be a direct transition between cavities $a_{1}$ and $a_{2}$,
and the transitions between the cavity modes and the mechanical mode are forbidden. Figure 3
shows the evolution of the populations of the system. Compared with the Fig. 1(b), the population transfer of the cavity mode $a_{1}$ and $a_{2}$ can be achieved within a short time.

\begin{figure}[t]
\includegraphics*[ width=4.2cm, height=2.5cm]{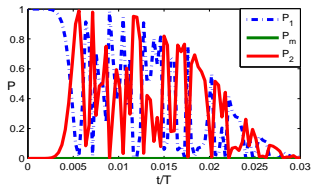}
\caption{(color online)  Simulation of the quantum state populations transfer of the cavity mode $a_{1}$ (dashed blue line), cavity mode $a_{2}$ (solid red line), and  mechanical mode $b$ (solid green line).}
\end{figure}

\section{discussion and conclusion}
When the transitionless quantum driving was employed to speed up adiabatic state conversion. It is important meaning to investigating the influence of energetic cost.
The cost of counterdiabatic driving can be written in general as \cite{23, 24, 25, 26}
\begin{eqnarray}
C=\frac{1}{T}\int_{0}^{T}||H(t)||dt,
\end{eqnarray}
with the Frobenius norm $||A||=\sqrt{\texttt{Tr}[A^{\dag}A]}$, where $A$ is an operator. For the Hamiltonian $\mathcal{N}+\mathcal{H}_{I}$, we obtain
\begin{eqnarray}
C=\frac{1}{T}\int_{0}^{T}\sqrt{\sum_{m}[E^{2}_{m}+\mu_{m}(t)]}dt,
\end{eqnarray}
where $E_{m}$ are the eigenvalues of the Eq. (\ref{b}) and $\mu_{m}(t)=\langle\partial_{t}\lambda_{m}(t)|\partial_{t}\lambda_{m}(t)\rangle
-|\langle\lambda_{m}(t)|\partial_{t}\lambda_{m}(t)\rangle|^{2}$. The instantaneous cost is
\begin{eqnarray}
\partial_{t}C=\sqrt{2g_{0}^{2}-3\kappa^{2}/4+2\vartheta^{2}}.
\end{eqnarray}
This equation shows an increase in the instantaneous cost to speed up adiabatic
evolutions compared to their adiabatic counterparts. Figure (4) illustrates the behavior of $\partial_{t}C/g_{0}$ versus the $g_{0}$ for different $\vartheta$ with the $\kappa=0.01g_{0}$.
It is apparent that the $\partial_{t}C/g_{0}$ inversely proportional to the increasing $g_{0}$. And the
speeding up adiabatic cost recover the cost of adiabatic with the increasing $g_{0}$.
\begin{figure}[t]
\includegraphics*[ width=4.2cm, height=2.8cm]{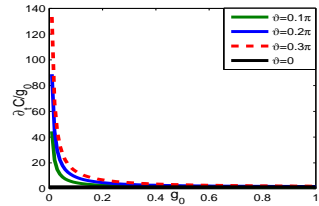}
\caption{(color online)  The $\partial_{t}C/g_{0}$ versus the $g_{0}$ for different $\vartheta$, $\vartheta=0$ (black line),$\vartheta=0.1\pi$ (green line), $\vartheta=0.2\pi$ (blue line), and $\vartheta=0.3\pi$ (dashed red line).}
\end{figure}

The real values of the experimental parameters are given to show the feasibility
of our scheme \cite{22}. The mechanical mode has the resonance frequency $\omega_{m}=2\pi\times3.68$GHz with an intrinsic damping rate $\gamma=2\pi\times35$kHz. The optomechanical coupling strength $g_{i}=2\pi\times910$kHz. The cavity mode has the resonance frequency $\omega_{0}=2\pi\times195$THz. It is clear that the parameters used in our scheme has the same order of magnitude with the actual experimental implementation.

In summary, we have proposed a scheme to speed up adiabatic state conversion in optomechanical system. The perfect population transfer between cavity mode $a_{1}$ and $a_{2}$ can be implemented. Also, this scheme has potential application in generating entanglement and realizing quantum logic gates.
\begin{acknowledgments}
 FYZ was supported by the National Science Foundation of China under Grant Nos. [11505024, 11447135, 11505023, and 11605024] and the Fundamental Research Funds for the Central Universities No. DC201502080407. YX was supported by the National Natural Science Foundation of China under Grants No. 11575045, No. 11374054 and No. 11674060,
and the Major State Basic Research Development Program of China
under Grant No. 2012CB921601.
\end{acknowledgments}

\end{document}